\documentclass[structabstract]{aa}
\usepackage{graphicx}
\usepackage{txfonts}
\usepackage{natbib}
\begin{document}
   \title{Spacecraft VLBI and Doppler tracking: algorithms and implementation}
   \titlerunning{S/C VLBI \& Doppler tracking: algorithms and implementation}

   \author{D.A.~Duev
          \inst{1,2}
          \and
          G.~Molera Calv\'{e}s
          \inst{1,3}
          \and
          S.V.~Pogrebenko
          \inst{1}
          \and
          L.I.~Gurvits
          \inst{1,4}
          \and
          G.~Cim\'{o}
          \inst{1}
          \and
          T.~Bocanegra~Bahamon
          \inst{1,4,5}
          }
   \institute{Joint Institute for VLBI in Europe, P.O. Box 2, 7990 AA Dwingeloo,
              The Netherlands\\
              \email{duev@jive.nl, molera@jive.nl, pogrebenko@jive.nl, lgurvits@jive.nl, cimo@jive.nl,
                         bocanegra@jive.nl}
         \and
             Faculty of Physics, Lomonosov Moscow State University,
             GSP-1, Leninskie gory, 119991 Moscow, Russia
         \and
            Aalto University Mets\"{a}hovi radio observatory, Mets\"{a}hovintie 114, Kylm\"{a}l\"{a}, FIN-02540,
            Finland
        \and
            Department of Astrodynamics and Space Missions,
            Delft University of Technology, 2629 HS Delft, The Netherlands
        \and
            Shanghai Astronomical Observatory,
            80 Nandan Road, Shanghai 200030, China
             }

   \date{Received 25 January 2012 / Accepted 9 March 2012}


  \abstract
   {}
   {We present the results of several multi-station Very Long Baseline Interferometry (VLBI) experiments conducted with the ESA spacecraft Venus Express as a target. To determine the true capabilities of VLBI tracking for future planetary missions in the solar system, it is necessary to demonstrate the accuracy of the method for existing operational spacecraft.}
   {We describe the software pipeline for the processing of phase referencing near-field VLBI observations and present results of the ESA \object{Venus Express} spacecraft observing campaign conducted in 2010-2011.}
   {We show that a highly accurate determination of spacecraft state-vectors is achievable with our method. The consistency of the positions indicates that an internal rms accuracy of 0.1 mas has been achieved. However, systematic effects produce offsets up to 1 mas, but can be reduced by better modelling of the troposphere and ionosphere and closer target-calibrator configurations.}
   {}


   \maketitle
%

\section{Introduction}
The VLBI spacecraft tracking technique has been successfully applied to a number of deep-space missions. The technique imposes minimal requirements on the mission's on-board instrumentation and can operate with almost any radio signal from a spacecraft, provided it is sufficiently powerful and phase-stable. Among VLBI-tracked missions one can mention the VEGA Venus atmosphere balloons \citep{Vega}, the Huygens Probe during its descent to the surface of Saturn's moon Titan \citep{Lebreton}, VLBI tracking with the European VLBI Network (EVN) antennas of the controlled impact of ESA's Smart-1 probe on the surface of the Moon \citep{Smart}, VLBI tracking of NASA's Mars Exploration Rover B spacecraft during its final cruise phase \citep{Lanyi}, VLBI tracking of the Cassini spacecraft at Saturn \citep{Jones}, and the latest results achieved with ESA's Venus Express (VEX) VLBI spacecraft observations and the ESA Mars Express (MEX) Phobos-flyby \citep{VEXMEX} monitored by the EVN radio telescopes.

The Planetary Radio Interferometry and Doppler Experiment (PRIDE) is an initiative by the Joint Institute for VLBI in Europe (JIVE). Designed as a multi-purpose, multi-disciplinary enhancement of planetary missions science return, PRIDE is able to provide ultra-precise estimates of spacecraft state vectors based on the phase-referenced VLBI tracking and radial Doppler measurements. These can be used for a variety of scientific applications including planetary science (measurements of tidal deformations of planetary moons, atmosphere dynamics and climatology as well as seismology, tectonics, internal structure and composition of planetary bodies), ultra-precise celestial mechanics of planetary systems, gravimetry and fundamental physics (e.g. tests of General Relativity or study of the anomalous accelerations of deep-space probes \citep{Turyshev}). Spacecraft Doppler tracking is known to be the only possible way of detecting low-frequency ($10^{-5}-1$ Hz) gravitational waves \citep{Kopeikin}. Studies of interplanetary scintillations (Molera Calv\'{e}s et al., in prep.) are among the potentially rewarding by-products of PRIDE. PRIDE can also provide diagnostics of deep-space missions (``health check") and direct-to-Earth delivery of a limited amount of critical data (e.g. Friedman et al.\citet{Fridman}).

PRIDE can be conducted with literally any radio-emitting spacecraft. Current PRIDE 'customers' include ESA's VEX and MEX spacecraft and the Russian Space Agency's Space VLBI mission RadioAstron \citet{Radioastron}.
PRIDE has been considered in a number of design studies, such as EVE \citet{PRIDE-EVE}, JUICE/Laplace \citet{JUICE-1,JUICE-2}, TandEM \citet{TandEM}, Kronos \citet{Kronos}, and several others.

In this paper, we describe the principles and pipeline of the data acquisition, processing and analysis within the scope of PRIDE and present the results of the first large-scale observational EVN campaign conducted in 2009--2011 with ESA's VEX spacecraft.

\section{Observations and data processing pipeline}
In phase--referencing VLBI, the relative position of a specific source is determined by observing a known source nearby (i.e. at a short angular distance on the celestial sphere), called a phase-reference calibrator, and applying calibration values of its fringe phase to the fringe phases of a target source. The most accurate measurements can be achieved when both the target and calibrator sources are within the same primary beam of a VLBI antenna or no more than a few degrees apart \citep{Ros}. VLBI phase referencing for spacecraft tracking involves alternated observations of the radio signal from the spacecraft and a nearby calibrator, most commonly a quasar. The best switching time between the calibrator and target sources for out-of-beam phase-referencing ranges from tens of seconds to several minutes. In the out-of-beam regime, the ideal scan duration on each source depends primarily on the efficiency of the radio telescopes, their system temperature, and the slewing speed. In the observations described below, the scan duration was set to 2--5 minutes.

Spacecraft VLBI-tracking data processing and analysis tools are based on two major software engines: the ultra-high spectral resolution and phase detection software package SWSpec/SCTrack/Phase-Lock-Loop, developed at the  Mets\"{a}hovi Radio Observatory in collaboration with JIVE\footnote{Wagner,~J., Molera Calv\'{e}s,~G. and Pogrebenko,~S.V. 2009-2012, Mets\"{a}hovi Software Spectrometer and Spacecraft Tracking tools, Software Release, GNU GPL, http://www.metsahovi.fi/en/vlbi/spec/index}, and the EVN Software Correlator at JIVE SFXC \citep{SFXC}.

Successful detection, processing, and analysis of spacecraft VLBI-tracking data requires usage of a near-field delay model and computation of near-field analogues of baseline {\it uv}-projections. To address these tasks, a software package VINT (VLBI in the near-filed toolkit) was developed at JIVE and is described in detail below.

\subsection{Observations}
{\bf Preparation for observations, \ scheduling.}
In standard VLBI observations, coordinates of a source are considered to be constant during the observing run for all participating telescopes. Usually, the source position in the geocentric equatorial inertial coordinate system J2000.0 is used. For near-field VLBI observations of spacecraft, the target can move rapidly across the primary beam of a telescope. Therefore, the position of a spacecraft is calculated at different epochs and is then used for pointing the
antenna at those particular epochs. In the ``ultra" near-field case, i.e. when the target is within (or very close to) the synthesised aperture of a VLBI array\footnote{this is true for VLBI observations of, e.g., GNSS satellites or the Russian Space Agency's orbital radio telescope RadioAstron}, so-called ``cross-eyed" scheduling should be used. This means that the spacecraft J2000.0 positions are different in the schedules for different VLBI stations.

{\bf Raw data.}
The down-converted signals observed by radio telescopes are recorded with the standard VLBI recording systems Mark5 A/B \citep{Mk5}, or the PC-EVN \citep{PCEVN}. Both systems were designed exclusively to record formatted data of astronomic and geodetic VLBI observations, allowing high data-rate ranging up to 2 Gbps.

   \begin{figure}
   \centering
   \includegraphics[width=240pt,clip]{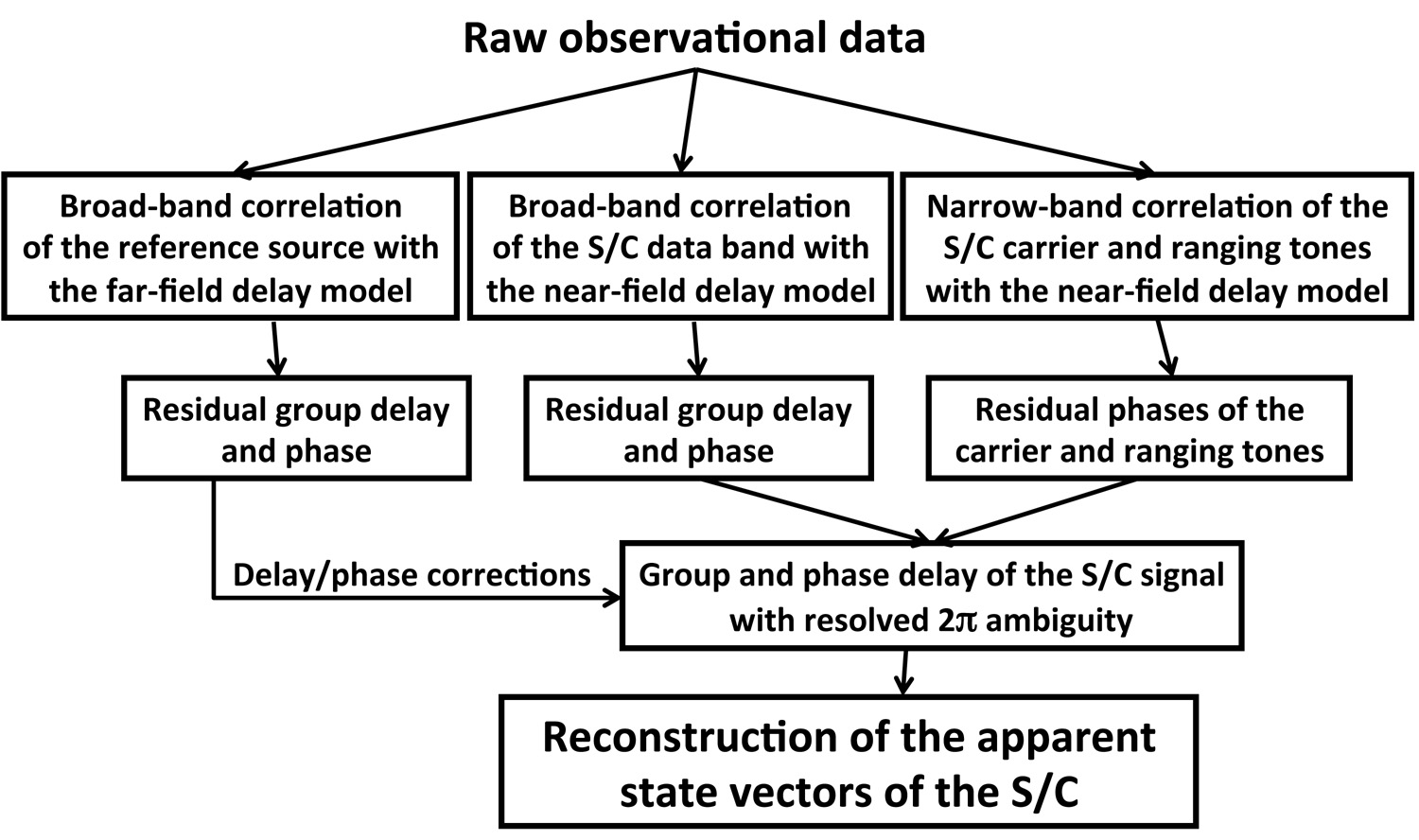}
   \caption{Data processing pipeline of phase-referencing VLBI observations of spacecraft.}
   \label{pipeline}
    \end{figure}

Fig. \ref{pipeline} represents the data processing pipeline that is described in detail below.

\subsection{Broadband correlation of reference-source signal and S/C data band}

{\bf Initial processing.}
Once at JIVE, the raw data are correlated with the JIVE Software Correlator SFXC \citep{SFXC}, which has been developed at JIVE and is currently used as a production correlator for some EVN observations. The SFXC is based on the original design developed for VLBI tracking of the Huygens Probe \citep{Huygens} and is capable of  supporting both the far-field and near-field delay models and processing data in various ``flavours" of the Mk5 format.

When correlated with the SFXC, the fringe finder\footnote{a strong compact source observed to find interferometric fringes} data are used for the clock search and bandpass correction. Also, a two-way Doppler phase correction (obtained with the SCTrack software, see next subsection) is applied to the spacecraft data to avoid frequency smearing.

The correlation results are saved in a measurement set format \citep{ms}. After this, the amplitudes and phases of the cross-correlation function are extracted for further processing.

{\bf Group delay extraction.}
The complex cross-correlation functions (CCF) corrected for clock/clock-rate offsets are smoothed in the frequency domain. After this, the phase is extracted for each averaged spectrum and fitted by a linear function. A group delay for the centre of each averaging interval is derived as the slope of the function $\tau_\mathrm{gr} = \mathrm{d} \phi_\mathrm{fit} / \mathrm{d} f$, where $f$ denotes frequency.

\subsection{Narrow--band correlation of S/C signal}

{\bf SWSpec, SCTrack and phase-lock loop.}
Narrow band correlation of the spacecraft signal is performed with the SWSpec/SCTrack software package described in Molera Calv\'{e}s et al. \cite{GuifreIPPW}.

The high-resolution spectrometer software (SWspec) allows the initial detection of the carrier tone of the spacecraft and temporal evolution of its frequency over the entire scan. In each SWspec pass, one of the raw data channels is extracted from the input data file. The software performs an accurate window-overlapped discrete Fourier transform (DFT) and time integration of the spectrum. The generated time-integrated spectra are processed to determine the moving phase of the S/C tone frequencies along the scan. The series of frequency detections of the carrier line are fitted to a sixth-order phase-stopping polynomial.

The core of the spacecraft tone tracking software (SCtracker) performs a phase polynomial correction to compensate for the Doppler shift, single and/or multi-tone tracking, signal filtering for each tone and, finally, phase detection of the carrier line with respect to the local H-maser clock.

The final post-analysis of the narrow-bands is performed with the phase-lock loop (PLL) software. The PLL runs high-precision iterations of the phase polynomial fit on the filtered complex narrow-band signals. The implementation of the PLL allows filtering down the tones as often as desired to achieve mHz accuracy around the spacecraft carrier line or other tones in the band. The final residual phase in a specific stopped band is determined with respect to the set of subsequent frequency/phase polynomials initially applied.

{\bf Reduction of phases to a common phase centre.}
Topocentric phase detections for each station are reduced to the common phase centre (usually the geocentre) by applying the geocentric signal delay, $\tau_\mathrm{gc}$, for the near-field case (detailed description of the model used is given in the next subsection):
\begin{equation}
\phi_\mathrm{gc}(t) = \phi_\mathrm{tc}(t-\tau_\mathrm{gc}).
\label{eq1}
\end{equation}
These geocentric phases are used as Doppler phase corrections for the broad-band correlation with SFXC.
Finally, for each baseline {\it i-j}, differential phases $\phi_{ij} = \phi_{j, \mathrm{gc}} - \phi_{i, \mathrm{gc}}$
are formed.

\subsection{Resolution of the $2\pi$-ambiguity}
A straightforward correlation of the wide-band spacecraft signal allows deriving the group delay $\tau_\mathrm{gr} = \mathrm{d} \phi / \mathrm{d} f$, while the narrow-band correlation of the S/C carrier tone allows deriving the phase delay $\tau_\mathrm{ph} = \phi / f_0$, where $f_0$ is the observational frequency. These are two independent measurables, which should give the same phase solution. The rms error of the group delay is $\sigma(\tau_\mathrm{gr}) = 1/(\Delta F \times S/N)$, with $\Delta F$ being the bandwidth of, e.g. 8 or 16 MHz for the calibrator and about 1 MHz for the S/C but with an uneven distribution of S/N (signal-to-noise ratio) over the signal band. The rms error of the phase delay for a narrow-band signal is given by $\sigma(\tau_\mathrm{ph}) = 1/(f_0 \times \sqrt{S/N})$, where the S/N level for the spectral line is very high, so that even the $\sqrt{S/N}$ may be larger than the S/N for the broad band case. In addition to this, $f_0 \gg \Delta F$, so that the rms error of group delay is significantly larger than that of the phase delay. This in turn results in a different scatter of the phase solution resulting from group and phase delays. 

The group delay, although not that accurate, is a ``true delay", while the phase delay is derived from the measured phase and may be subject to a $2\pi$ ambiguity. Since phases are $2\pi$ uncertain, the phase delay may have a bias of several cycles of $\sim120$ ps for observations at X-band. Checking the phase delay versus group delay consistency by computing the uncertainty of the fit of the phase delay to the group delay and “shifting” the former by $n \cdotp 2\pi$ (where $n$ is an integer number) to obtain the best fit to the latter allows one to resolve the $2\pi$ ambiguity.

\subsection{Delay model}
The VLBI delay model is formulated in the barycentric celestial reference system (BCRS). This means that the delay is computed in the time scale of the BCRS and is subsequently transformed to the time--scale used for timing the measurements on Earth. Thus several time--scale and station coordinate transformations are required.

The models described below are coded in Matlab and aggregated within the VINT software package.

{\bf Geocentric terrestrial to celestial frame transformation.}
As a first step, a transformation of station coordinates from the terrestrial reference system (realised by the international terrestrial reference frame, ITRF) to the geocentric celestial reference system (GCRS) is conducted according to the latest recommendations of the international Earth rotation service IERS 2010 conventions \citep{tn36} within the framework of the IAU 2000/2006 resolutions.

Station positions and velocities in the ITRF at the epoch $t_0$ (e.g. 1997.0 or 2000.0, depending on the realisation used), are reduced to the epoch of observation, taking the tectonic plate motion into account. A simple linear model is used:
\begin{equation}
{\bf r}_\mathrm{ITRF} = {\bf r}_\mathrm{0} + {\bf \dot{r}}_\mathrm{0} \cdotp (t-t_\mathrm{0}).
\end{equation}

At the next step, the station position, velocity, and acceleration in the GCRS are computed as follows:
\begin{eqnarray}
{\bf r}_\mathrm{GCRS} = &&T(t) \cdotp {\bf r}_\mathrm{ITRF}, \nonumber\\
&& ({\bf r}_\mathrm{GCRS} \rightarrow {\bf \dot{r}}_\mathrm{GCRS}, {\bf \ddot{r}}_\mathrm{GCRS}; \ T(t) \rightarrow
\dot{T}(t), \ddot{T}(t) ),
\end{eqnarray}
where the transformation matrix $T(t) = Q(t) \cdotp R(t) \cdotp W(t)$ with $Q(t)$, $R(t)$ and $W(t)$ being $3\times3$ matrices representing the motion of the celestial pole in the GCRS, rotation of the Earth around the axis associated with the pole and polar motion, correspondingly, and the derivatives $\dot{T}(t), \ddot{T}(t)$ are calculated numerically.

{\bf Displacements of station reference points.}
Station positions in the GCRS should be subsequently corrected for various geophysical effects. These include displacements caused by the solid Earth tides arising from the direct effect of the external tide-generating potential, ocean tidal loading, diurnal and semidiurnal atmospheric pressure loading, and the centrifugal perturbations caused by the pole tide-driven Earth rotation variations. These effects are taken into account following the IERS conventions 2010 \citep{tn36}.

{\bf Geocentric to Barycentric transformation.}
Finally, the Lorentz transformation needs to be applied to the station GCRS position ${\bf r}_\mathrm{gc}$, velocity ${\bf \dot{r}}_\mathrm{gc}$ and acceleration ${\bf \ddot{r}}_\mathrm{gc}$ to transform it in the solar system barycentric reference frame BCRS:

\begin{equation}
{\bf r}_\mathrm{bc} = \big(1 - L_C - \frac{\gamma U_\mathrm{E}}{c^2}\big) \cdot {\bf r}_\mathrm{gc} - \frac{1}{2c^2}( {\bf V}_\mathrm{E} \cdot {\bf r}_\mathrm{gc}) \cdot {\bf V}_\mathrm{E} + {\bf R}_\mathrm{E} 
\end{equation}
\begin{eqnarray}
{\bf \dot{r}}_\mathrm{bc} =&&\big(1 - \frac{(1+\gamma) U_\mathrm{E}}{c^2} - \frac{V_\mathrm{E}^2}{2c^2} - \frac{{\bf V}_\mathrm{E} \cdot {\bf \dot{r}}_\mathrm{gc}}{c^2}\big) \cdot {\bf \dot{r}}_\mathrm{gc}\nonumber\\
&&+ {\bf V}_\mathrm{E} \cdot \big( 1 - \frac{1}{2c^2}( {\bf V}_\mathrm{E} \cdot {\bf \dot{r}}_\mathrm{gc}) \big)
\end{eqnarray}
\begin{eqnarray}
{\bf \ddot{r}}_\mathrm{bc}=&&\big(1 + L_C - \frac{(2+\gamma) U_\mathrm{E}}{c^2} - \frac{V_\mathrm{E}^2}{c^2} - \frac{{2\bf V}_\mathrm{E} \cdot {\bf \dot{r}}_\mathrm{gc}}{c^2}\big) \cdot {\bf \ddot{r}}_\mathrm{gc}\nonumber\\
&&-\frac{1}{2c^2}( {\bf V}_\mathrm{E} \cdot {\bf \ddot{r}}_\mathrm{gc}) \cdot ({\bf V}_\mathrm{E} + 2{\bf
\dot{r}}_\mathrm{gc}) + {\bf A}_\mathrm{E},
\end{eqnarray}

where
$L_C = 1.48082686741 \times 10^{-8}, \ $
$U_\mathrm{E} = \sum\limits_{j \neq i} \frac{GM_j}{r_{\mathrm{E}j}} \ $ is the Newtonian potential of all solar system bodies excluding the Earth, evaluated at the geocenter, \
$\gamma $ is the PPN parameter (equal to 1 in General Relativity), \
${\bf R}_\mathrm{E}$, ${\bf V}_\mathrm{E}$ and ${\bf A}_\mathrm{E}$ are the position, velocity and acceleration of the Earth in the BCRS, respectively,\
and $c$ is the speed of light in a vacuum.

{\bf Time--scale transformations.} The time--scale used for measurement timing at stations is the coordinated universal time (UTC), which differs from the physically realised international atomic time scale, TAI, by a number of leap-seconds (currently, 34 seconds). The latter, in turn, differs by 32.184 seconds from the terrestrial time (TT), which is the theoretical time--scale for clocks at sea-level. The TT is a scaled version of the geocentric coordinate time (TCG, the time coordinate of the IAU space--time metric GCRS), which eliminates the combined effect on the terrestrial clock of the gravitational potential from the Earth and the observatory's diurnal speed in TCG \citep{tn36}:
\begin{equation}
TCG = TT+L_\mathrm{G} \cdotp (JD_\mathrm{TT} - TT_0),
\end{equation}
where $TT_0$ is TT at 1977 January 1.0 TAI, $JD_\mathrm{TT}$ is TT, both expressed as Julian date, and $L_\mathrm{G} =
6.969290134 \times 10^{-10}$.

The time--scale used in the ephemerides of planetary spacecraft, as well as that of solar system bodies, is the barycentric dynamical time (TDB), a scaled version of the barycentric coordinate time (TCB) - the time coordinate of the IAU space--time metric BCRS. The TDB stays close to TT on the average by suppressing a drift in TCB due to the combined effect of the terrestrial observer's orbital speed and the gravitational potential from the Sun and planets by applying a linear transformation \citep{tn36}
\begin{equation}
TDB = TCB - L_\mathrm{B} \cdotp (JD_\mathrm{TCB} - T_0)\cdotp 86400+TDB_0,
\end{equation}
where $JD_\mathrm{TCB}$ is TCB expressed as Julian date, $T_0$ = 2443144.5003725, $L_\mathrm{B} = 1.550519768 \times
10^{-8}$ and $TDB_0 = −6.55 \times 10^{-5}$ s.

The difference between TCG and TCB involves a transformation accounting for the orbital speed of the geocenter and the gravitational potential from the Sun and planets with the accuracy of $O(c^{-4})$ given by
\begin{equation}
TCB-TCG = c^{-2} \Big( \int_\mathrm{t_0}^{t} \big[ \frac{V_\mathrm{E}^2}{2} +U_\mathrm{E}({\bf R}_\mathrm{E}) \big]
\mathrm{d}t+ {\bf V}_\mathrm{E} \cdotp ({\bf R} - {\bf R}_\mathrm{E}) \Big),
\end{equation}
where ${\bf R}_\mathrm{E}$ denotes the BCRS position of the Earth and ${\bf R}$ is the BCRS position of the
observing station.

Computation of the difference TCB $-$ TCG is performed on the basis of the Fairhead-Bretagnon model \citep{Fairhead} in its full form, which yields on accuracy of a few nanoseconds. The full time--scale transformation chain is as follows:
$UTC \rightarrow TAI \rightarrow TT \rightarrow TCG \rightarrow TCB \rightarrow TDB$.

{\bf VLBI delay model for extragalactic sources.}
For calculating the VLBI signal delay for the extragalactic sources, we use the so-called consensus model recommended by the IERS \citep{tn36}, which is based on the plane-wave approximation when the distance to a source is assumed to be infinite.
In the BCRS, the vacuum delay equation is
\begin{equation}
T_2 - T_1 = -\frac{{\bf K}}{c} \cdotp ({\bf R}_2(T_2) - {\bf R}_1(T_1)) + \Delta T_\mathrm{grav}.
\end{equation}

The total barycentric vacuum delay is then converted into the geocentric one by applying the Lorentz transformations and is given by
\begin{eqnarray}
t_2-t_1=&&\bigg( 1+\frac{{\bf K} \cdotp ({\bf V}_\mathrm{E} + {\bf \dot{r}}_\mathrm{2, gc})}{c} \bigg) ^{-1} \cdotp
\bigg( \Delta T_\mathrm{grav} - \frac{{\bf K} \cdotp {\bf b}}{c} \cdotp \nonumber\\
&& \cdotp \bigg[ 1 - \frac{(1+\gamma) U_\mathrm{E}}{c^2} - \frac{V_\mathrm{E}^2}{2c^2} - \frac{{\bf V}_\mathrm{E} \cdot {\bf \dot{r}}_\mathrm{2, gc}}{c^2} \bigg] \nonumber\\
&& - \frac{{\bf V}_\mathrm{E} \cdotp {\bf b}}{c^2} \cdotp (1+ \frac{{\bf K} \cdotp {\bf V}_\mathrm{E}}{2c}) \bigg).
\end{eqnarray}

Here ${\bf K}$ is the unit vector from the barycentre to the source in the absence of gravitational or aberrational bending, \
${\bf b} = {\bf r}_2(t_1) -{\bf r}_1(t_1) $ is the GCRS baseline vector at the time of signal arrival at the first station. The general relativistic delay, $\Delta T_\mathrm{grav}$, is given for the $J_\mathrm{th}$ gravitating body by
\begin{equation}
\Delta T_{\mathrm{grav}, J} = 2\frac{GM_J}{c^3} \ln{\frac{ R_{1J}+ {\bf K} \cdotp {\bf R}_{1J}}{ R_{2J}+ {\bf K} \cdotp
{\bf R}_{2J}} }
\end{equation}

with ${\bf R}_{1J} = {\bf R}_{1J}(t_1) = {\bf R}_1(t_1) - {\bf R}_J(t_{1J})$ and ${\bf R}_{2J} = {\bf R}_{2J}(t_1) =
{\bf R}_2(t_1) - \frac{{\bf V}_\mathrm{E}}{c}({\bf K} \cdotp {\bf b}) - {\bf R}_J(t_{1J})$, where ${\bf R}_1$ and ${\bf
R}_2$ are the barycentric radius--vectors of the first and second receivers, $t_\mathrm{1}$ is the TCG time of arrival of the radio signal at the first VLBI receiver, $t_{1J}$ - the moment of closest approach of the photon to the gravitating body J in TCG. The total gravitational delay is the sum over all gravitating bodies including the Earth: $\Delta T_\mathrm{grav} = 􏰔 \sum\limits_J\Delta T_{\mathrm{grav},J}$ .

{\bf Delay model for VLBI observations of spacecraft.}

   \begin{figure}
   \centering
   \includegraphics[width=200pt,clip]{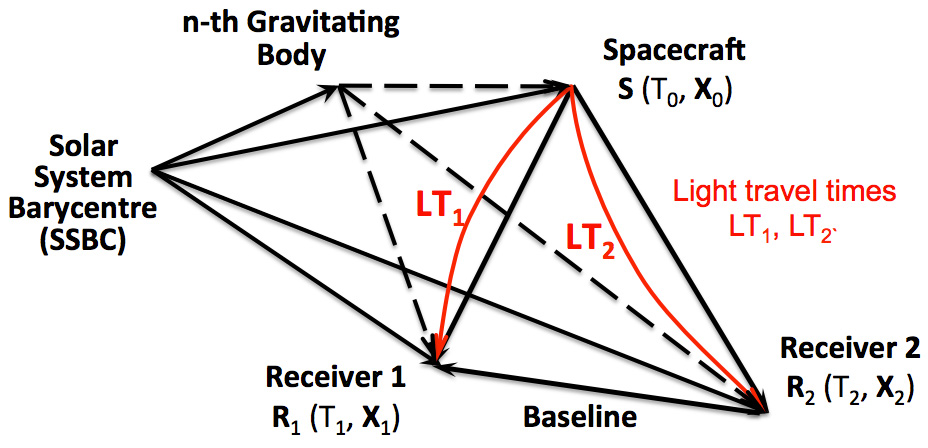}
   \caption{Geometry of VLBI observations of spacecraft in the Barycentric celestial reference frame.}
   \label{NF}
    \end{figure}

For spacecraft VLBI observations, a delay model should be able to account for the ``near--field" effects caused by finiteness of the source's distance. Indeed, the diffracted radio wave is considered to be in the near field if the distance to the source from the aperture $R\la D^2/\lambda$ \citep{BornWolf}, where $D$ is the characteristic size of the aperture and $\lambda$ is the wavelength. For a VLBI array with a synthesised aperture of $\sim 1000$ km observing at X-band ($\lambda\sim 3$ cm) this yields $R\lesssim 100$ AU. Note that near-field delays can differ from ``classic" ones by microseconds.

The signal delay in the TDB-frame is defined as the difference between two light travel times $LT_1$ and $LT_2$ from the spacecraft to the first and the second station of the baseline (Fig. \ref{NF}). Given the reception time $T_1$ in TDB at the first station, the transmission time $T_0$ is calculated by iteratively solving the light-time equation for the 'down-leg' {\it 0-1}:

\begin{equation}
\label{LT}
T_1 - T_0 = \frac{R_\mathrm{01}}{c} + RLT_\mathrm{01},
\end{equation}

where $RLT_\mathrm{01}$ is the relativistic term, which accounts for the effects of both special and general relativity
\citep{moyer}:
\begin{eqnarray}
RLT_\mathrm{01} = && \frac{(1+\gamma) \cdotp GM_\mathrm{S}}{c^3} \cdotp \ln \frac{R_0^\mathrm{S} + R_1^\mathrm{S} +
R_\mathrm{01}^\mathrm{S} + \frac{(1+\gamma) \cdotp GM_\mathrm{S}}{c^2}}{R_0^\mathrm{S} + R_1^\mathrm{S} -
R_\mathrm{01}^\mathrm{S} + \frac{(1+\gamma) \cdotp GM_\mathrm{S}}{c^2}} \nonumber\\
&& + \sum_{B=1}^{10} \frac{(1+\gamma) \cdotp GM_B}{c^3} \cdotp \ln \frac{R_0^B + R_1^B + R_\mathrm{01}^B}{R_0^B + R_1^B
- R_\mathrm{01}^B}.
\end{eqnarray}

Here $S$ denotes the Sun and $B$ runs over all solar system planets and the Moon, ${\bf R}_0$ is the spacecraft barycentric position\footnote{for the ESA spacecraft, the ephemerides are provided by the ESA European Space Operations Centre in Darmstadt, Germany (ESOC)},

\begin{equation}
{\bf R}_\mathrm{01} = {\bf R}_1(T_1) - {\bf R}_0(T_0)
\end{equation}

\begin{equation}
{\bf R}_i^{\alpha} = {\bf R}_i(T_i) - {\bf R}_\mathrm{\alpha}(T_i), \ i=0,1; \ \alpha=S, B
\end{equation}

\begin{equation}
{\bf R}_\mathrm{01}^{\alpha} = {\bf R}_1^{\alpha}(T_1) - {\bf R}_0^{\alpha}(T_0), \ \alpha=S, B.
\end{equation}

At each iteration step the correction to the $T_0$ estimate is given by
\begin{equation}
\Delta T_0 = \frac{T_1 - T_0 - R_\mathrm{01}/c - RLT_\mathrm{01}}{1 - \dot{p}_\mathrm{01}/c},
\end{equation}
where
\begin{equation}
\dot{p}_\mathrm{01} = \frac{ {\bf R}_\mathrm{01} }{R_\mathrm{01}} \cdotp \dot{{\bf R}}_0(T_0).
\end{equation}

Given the solution $T_0$ of equation (\ref{LT}), the light-time equation for the down-leg {\it 0-2} is solved for reception time $T_2$ in the same manner as described above.

The difference $T_2-T_1$ represents the VLBI delay in the barycentric TDB-frame, which is then transformed into the geocentric TT-frame:
\begin{equation}
t_2 - t_1 = \Big( \frac{T_2 - T_1}{1 - L_C} \cdotp \big[ 1 - \frac{1}{c^2} \big( \frac{V_\mathrm{E}^2}{2} + U_\mathrm{E}
\big) \big] - \frac{{\bf V}_\mathrm{E} \cdotp {\bf b}}{c^2} \Big) \cdotp \Big(1+\frac{{\bf V}_\mathrm{E} \cdotp
\dot{{\bf r}}_\mathrm{2,gc}}{c^2}\Big)^{-1}.
\end{equation}

The performance of the geometric delay model described here was checked by comparing it with one of the most accurate up-to-date analytical models by Sekido \& Fukushima \citet{SekidoFukushima}, which proved to be consistent at the picosecond level.

{\bf Propagation and other additional impacts on the signal delay.}
   \begin{figure}
   \centering
   \includegraphics[width=240pt,clip]{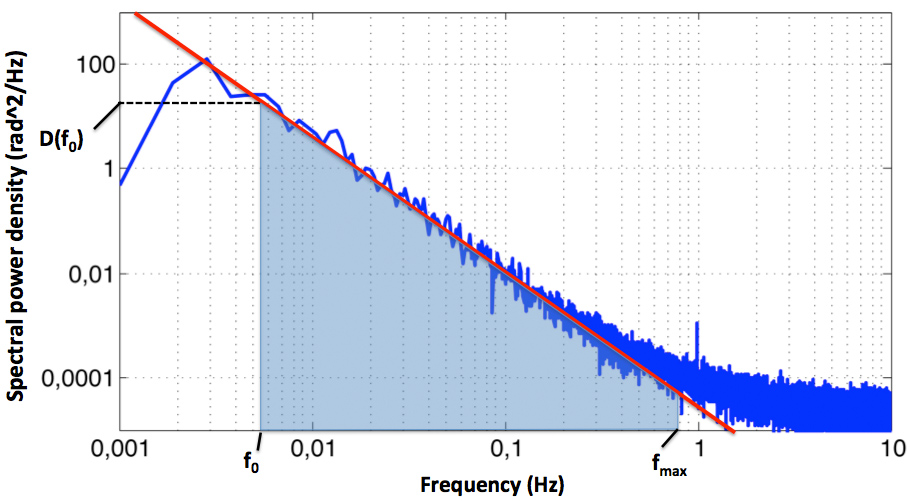}
   \caption{Kolmogorov spectrum of phase scintillations.}
   \label{scint}
    \end{figure}
Random fluctuations of a radio wave phase caused by propagation effects are dominated by the interplanetary scintillations in the solar wind. The power spectrum of these fluctuations is well-represented by the Kolmogorov spectrum \citep{Kolmogorov}. The scintillation phase is given by
\begin{equation}
\phi_\mathrm{sc} = \Big[ \int_\mathrm{f_0}^{f_\mathrm{max}} D(f_0) \cdotp \big( \frac{f}{f_0} \big)^{slope} \mathrm{d}f
\Big]^{1/2},
\end{equation}
where $D(f_0)$ is the spectral power density, the lower limit of integration $f_\mathrm{0}=1/\tau$ with $\tau$ being the length of the phase-referencing nodding cycle, the upper limit $f_\mathrm{max}$ is determined by the system noise level and $slope$ is the linear slope of the spectrum in logarithmic scale. For example, with a nodding cycle of $\tau = 180$~s, $f_0 = 1/\tau \approx 5.5$~mHz and the scintillation phase is proportional to the marked area in the Fig. \ref{scint}. Obviously, the shorter the nodding cycle, the better the target phase calibration.

The Earth's ionosphere and troposphere also induce an additional delay to the radio signal propagation. To mitigate the effect of the former medium, we use the vertical total electron content (vTEC) maps provided by the international GNSS service (IGS) on a daily basis with a two-hour temporal resolution on a global grid \citep{igs}. The IGS uses the single thin ionospheric layer model, therefore the vTEC should be properly mapped onto the direction of the target source to obtain the slant total electron content (TEC):

\begin{equation}
TEC = \frac{vTEC}{\cos{z'}},
\end{equation}

where $z'$ is the zenith distance of the target as seen from the model's ionospheric layer:
\begin{equation}
{z'} = \arcsin{\frac{R}{R+H} \cdotp \sin{z}},
\end{equation}
with $R$ being the mean radius of the Earth, $H$ the height of the ionospheric layer and $z$ the apparent vacuum zenith distance of the target.

The ionospheric delay for each station can then be computed as
\begin{equation}
\tau_\mathrm{iono} = \frac{5.308018 \cdotp TEC}{4\mathrm{\pi}^2 \cdotp f^2},
\end{equation}
where $f$ - is the observational frequency (e.g. 8.4~GHz for observations in X-band).

There are two options for mitigating the tropospheric effect on the signal propagation delay. They are based either on ``ready-to-use" empirical models or ray-tracing through the numerical weather models (NWM). The former provides the tropospheric delays in the direction of the local zenith and a mapping function to project the delay onto the direction to the source. We use the Vienna mapping functions VMF1 \citep{VMF}, recommended in the IERS conventions 2010.
The precision of the empirical models is sufficient for many cases, whereas we employ the latter option when an extreme accuracy is required, e.g. if the target sources is very low above the horizon and/or the angular distance between the spacecraft and the phase-referencing source is large. The latter was the case in the VEX tracking experiment EM081c, which we describe in the next section. We use the ray-tracing algorithm from Hobiger et al. \citep{Hobiger} with some modifications adopted from Duev et al. \citep{Duev}. The meteorological data from the European centre for medium--range weather forecasts (ECMWF) are used.

Additionally, contributions to the delay due to the antenna axis offsets and smaller size effects due to thermal \citep{Nothnagel} and gravitational \citep{Sarti, ClarkThomsen} deformations of telescopes are also taken into account in our delay model.

\subsection{Estimation of corrections to the S/C state vector}

The main task of VLBI tracking of spacecraft is estimating the state vector as a function of time. This task has some similarities with traditional astrometric applications of the VLBI technique. However, because a spacecraft is in the near--field of the VLBI-synthesised aperture, several modifications of traditional approaches and formalisms of far--field VLBI are necessary.

{\bf Analogues of {\it uv}--coverage for near--field VLBI.}
The {\it uv}--projections of baselines used in the standard far--field VLBI are basically the degenerated (in the limit
of an infinite distance to the source) partial derivatives of the predicted geocentric delays $\tau_{ij}$ by the source
position lateral to the line of sight:
\begin{eqnarray}
u_{ij}=\partial_\mathrm{\varphi}\tau_{ij}\cdotp c, \
v_{ij}=\partial_\mathrm{\theta}\tau_{ij}\cdotp c, \
\end{eqnarray}

where $c$ is the speed of light in a vacuum and $\varphi$ and $\theta$ are the geocentric spherical angular coordinates of the source.

For near--field VLBI observations at the epoch $t$ the {\it uv}-projections of the baselines \emph{i-j} ($i=\overline{1,N-1}$, $j=\overline{i+1,N}$,  where $N$ is the number of participating stations), because of the finiteness of the distance to the source, should be replaced with the corresponding elements of the Jacobian $J$:
\begin{equation}
J \big| _{t} = \left( \begin{array}{cc}
\frac{\partial \tau_\mathrm{12}}{\partial \varphi} & \frac{\partial \tau_\mathrm{12}}{\partial \theta} \\
\vdots & \vdots \\
\frac{\partial \tau_{1N}}{\partial \varphi} & \frac{\partial \tau_{1N}}{\partial \theta} \\
\vdots & \vdots \\
\frac{\partial \tau_{N-1,N}}{\partial \varphi} & \frac{\partial \tau_{N-1,N}}{\partial \theta} \\
\end{array} \right) = \left( \begin{array}{cc}
\frac{\partial (\tau_1-\tau_2)}{\partial \varphi} & \frac{\partial (\tau_1-\tau_2)}{\partial \theta} \\
\vdots & \vdots \\
\frac{\partial (\tau_1-\tau_N)}{\partial \varphi} & \frac{\partial (\tau_1-\tau_N)}{\partial \theta} \\
\vdots & \vdots \\
\frac{\partial (\tau_{N-1}-\tau_N)}{\partial \varphi} & \frac{\partial (\tau_{N-1}-\tau_N)}{\partial \theta} \\
\end{array} \right),
\end{equation}

where $\tau_i$ is the geocentric delay for the station \emph{i} at $t$ and the partials in the Jacobian are computed numerically. In other words, the traditional {\it uv}-formalism is merely a special case (a target source at an infinitely large distance) of a more general formalism, the Jacobians.

{\bf Spacecraft imaging and state vector estimation.}

The ('dirty') image of the spacecraft at the epoch $t$ is recovered by performing the Fourier transform of the sampled ``visibility" function \citep[see][]{TMS}
\begin{equation}
I_{t}(l,m) = \int{ \Re { \big( S(u,v) \cdotp V(u,v) \cdotp e^{- 2\mathrm{\pi i} \cdotp (ul+vm)} \big) \ \mathrm{d}u \
\mathrm{d}v } } \ \Big|_t ,
\end{equation}

with the sampling function being the sum over all baselines of two--dimensional delta functions:
\begin{equation}
S(u,v) \ \Big|_t = \sum_{i=1}^{N-1}\sum_{j=i+1}^{N} \delta (u-u_{ij}, v-v_{ij}) \ \Big|_t
\end{equation}
and the normalised weighted 'visibility' function given by
\begin{equation}
V(u_{ij},v_{ij}) \ \Big|_t = w_{ij} \cdotp e^{-2\mathrm{\pi i} \cdotp \phi_{ij}} \ \Big|_t,
\end{equation}
where $w_{ij}$ is the weight coefficient for the baseline {\it i-j} at time $t$.

For a given time range $t=[t_s, t_e]$, the resulting image is reconstructed as a sum over all epochs within the time range:
\begin{eqnarray}
I(l,m) = && \sum_{t} I_t(l,m) \nonumber\\
&&= \sum_{t=t_s}^{t_e} \bigg( \sum_{i=1}^{N-1} \sum_{j=i+1}^{N} w_{ij} \cdotp \Re { \big( \mathrm{e}^{\mathrm{i} \cdotp 2\mathrm{\pi} \cdotp (u_{ij}l+v_{ij}m - \phi_{ij} )} \big) } \bigg) \bigg|_t.
\end{eqnarray}

The measurement equation for the epoch $t$ is
\begin{equation}
\label{ME}
\overrightarrow{\Delta\phi} \ \big|_t = \big( J_{ij} \cdotp \overrightarrow{\Delta \alpha} \big) \ \big|_t,
\end{equation}
where $\overrightarrow{\Delta\phi}$ is a $[N_\mathrm{b} \times 1]$ vector of differential phases on all baselines
($N_\mathrm{b}=N\cdotp (N-1)/2$ is the number of baselines) and $\overrightarrow{\Delta \alpha}$ is the vector of
corrections to the a priori geocentric angular position of the spacecraft:
\begin{equation}
\overrightarrow{\Delta\phi} = \left( \begin{array}{c}
\phi_\mathrm{12} \\
\vdots \\
\phi_{1N} \\
\vdots \\
\phi_{N-1,N} \\
\end{array} \right), \
\overrightarrow{\Delta \alpha} = \left( \begin{array}{c}
\Delta\varphi \\
\Delta\theta
\end{array} \right).
\end{equation}

Finally, the astrometric solution of equation (\ref{ME}) in the least--squares sense for each epoch $t$ is given by
\begin{equation}
\overrightarrow{\Delta\alpha} \ \Big|_t = \Big( (J^\mathrm{T} \cdotp J)^{-1} \cdotp J^\mathrm{T} \cdotp
\overrightarrow{\Delta\phi} \Big) \ \Big|_t .
\label{LSQ}
\end{equation}

The main purpose for the introduction of the Jacobian formalism is to use it for the estimation of the corrections to the a priori position of a near-field VLBI target according to equations (\ref{ME})-(\ref{LSQ}). Imaging, although it can be used for the same purpose yielding the same result, is merely a ``by-product" of such a formalism, and is used in our paper mostly for the purposes of illustration. In addition to this, the closer the target is to an observer, the more significant the difference between the Jacobians and ``normal" {\it uv}-projections becomes.

\section{VEX VLBI observations and results}

   \begin{figure}
   \centering
   \includegraphics[width=250pt]{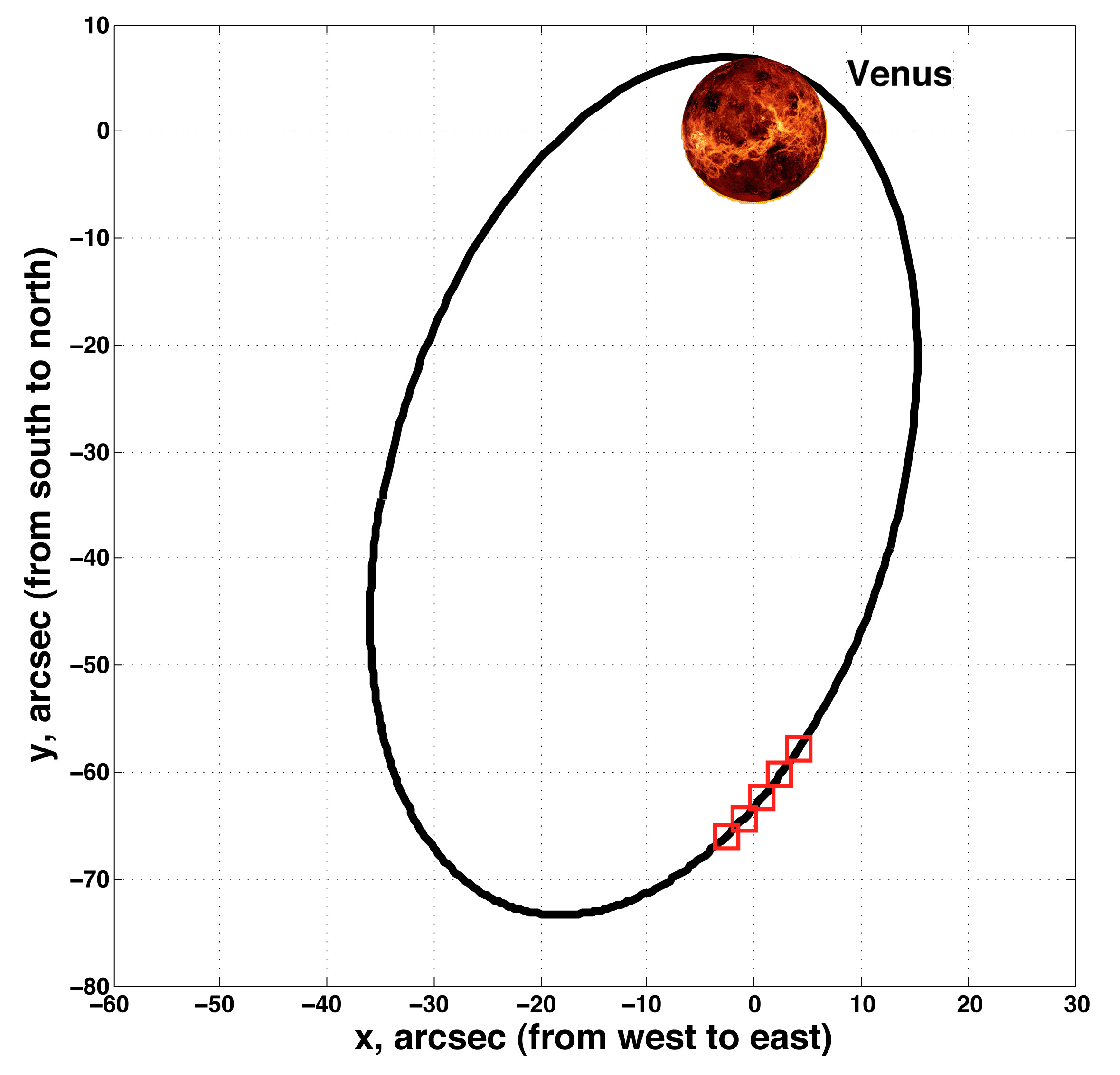}
   \caption{VEX orbit around Venus on 2011 March 28, as seen from the geocentre. Red boxes represent the position of VEX at the epochs when imaging was performed.}
   \label{VEXVenus}
    \end{figure}
%

   \begin{figure}
   \centering
   \includegraphics[width=260pt]{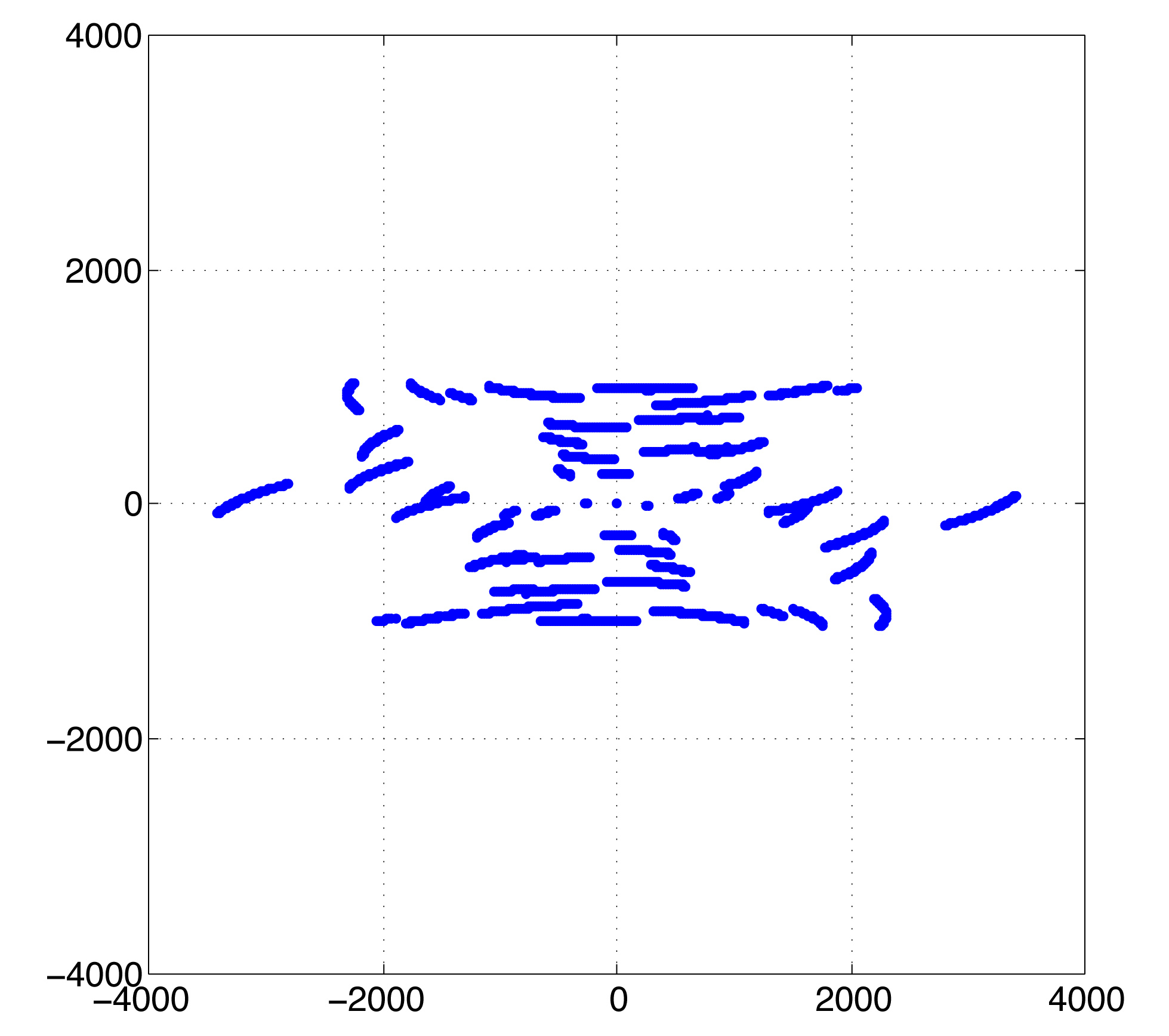}
   \caption{Near--field Jacobians for the VEX pointings in kilometers, 2011 March 28, the EVN project EM081. The longest east-west baseline -- Ys-Zc, the longest north-south baseline -- Mh-Ys}
   \label{UV}
    \end{figure}

ESA's Venus Express Spacecraft has been observed with the European VLBI Network (EVN) within the framework of the PRIDE initiative under the EVN observing project EM081.

From four requested observing epochs we used only three. The first two observing runs, on 2010 August 23 (2 hours long, 3 stations) and on 2010 September 20 (3 hours long, 2 stations) were mainly devoted to tuning-up and verification of the technique, namely developing and testing the S/C observation-related scheduling, data acquisition, transfer and processing methods with the near--field delay model.

Here we report the results obtained with the third observational run, which has been the most intensive experiment so far in terms of the number of participating telescopes: the EVN stations Onsala (On, Sweden), Wettzell (Wz, Germany), Medicina (Mc, Italy), Matera (Ma, Italy), Yebes (Ys, Spain), Mets\"{a}hovi (Mh, Finland), Svetloe (Sv, Russia), Zelenchukskaya (Zc, Russia), Hartebeesthoek (Hh, South Africa), and the NRAO Very Long Baseline Array (VLBA) station St.~Croix (Sc, USA).

The observations were conducted on 2011 March 28, from 08:45 to 11:30 UTC. The source \object{J2225-0457} (also known as \object{3C446})  was used as a fringe finder and for the clock search. The source \object{J2211-1328} with J2000.0 coordinates of $22^\mathrm{h}11^\mathrm{m}24\fs0994590, -13^{\circ}28\arcmin09.723950\arcsec$ \citep{LPetrov} was used as a phase--reference calibrator. The estimated coordinate error for this source is 0.11 mas in Right Ascension and 0.12 mas in Declination \citep{LPetrov}. All involved EVN radio telescopes were configured to observe the spacecraft signal in a phase-referencing mode, switching between the S/C and the phase reference source with a nodding cycle of 4 minutes, including a 20 second gap for re-pointing and antenna calibration. At On, Mc, Ma, Mh, Hh, we recorded 79 scans per station, including 2 for the fringe finder, 39 for the calibrator source and 38 for VEX. The stations Ys and Sc joined in starting from the $15^{th}$ scan owing to technical problems and limited visibility, respectively. The eastern--most stations Sv and Zc stopped observing according to schedule owing to the target sources' setting. During the observing run, the VEX spacecraft had mean coordinates $22^\mathrm{h}14^\mathrm{m}13^\mathrm{s}, -11^{\circ}41\arcmin22\arcsec$. At the time of the observations, Venus was at a solar elongation of 36 degrees and at a distance from Earth of 1.24~AU. Fig.~\ref{VEXVenus} shows the orbital position of VEX around Venus during the observations as seen from the geocentre. Fig.~\ref{UV} shows the pseudo-{\it uv}-coverage (Jacobians) of the EM081c VLBI array.

It should be pointed out that the observational setup for the experiment was not ideal -- target declinations ranging from $-11^{\circ}$ to $-13^{\circ}$ are unfavourable for the EVN array. In addition to this, the angular separation of $\sim2.5^{\circ}$ between Venus and the calibrator was at the edge of the isoplanatic patch\footnote{ Isoplanatic patch is an angular region on the sky within which the measured phases of a calibrator can be coherently applied to the source phases. Its size is determined by the atmospheric turbulence.} at low elevations caused by a low declination of the target, which lead to a relatively large unmodelled tropospheric/ionospheric phase difference between the calibrator and VEX.

   \begin{figure}
   \centering
   \includegraphics[width=260pt,clip]{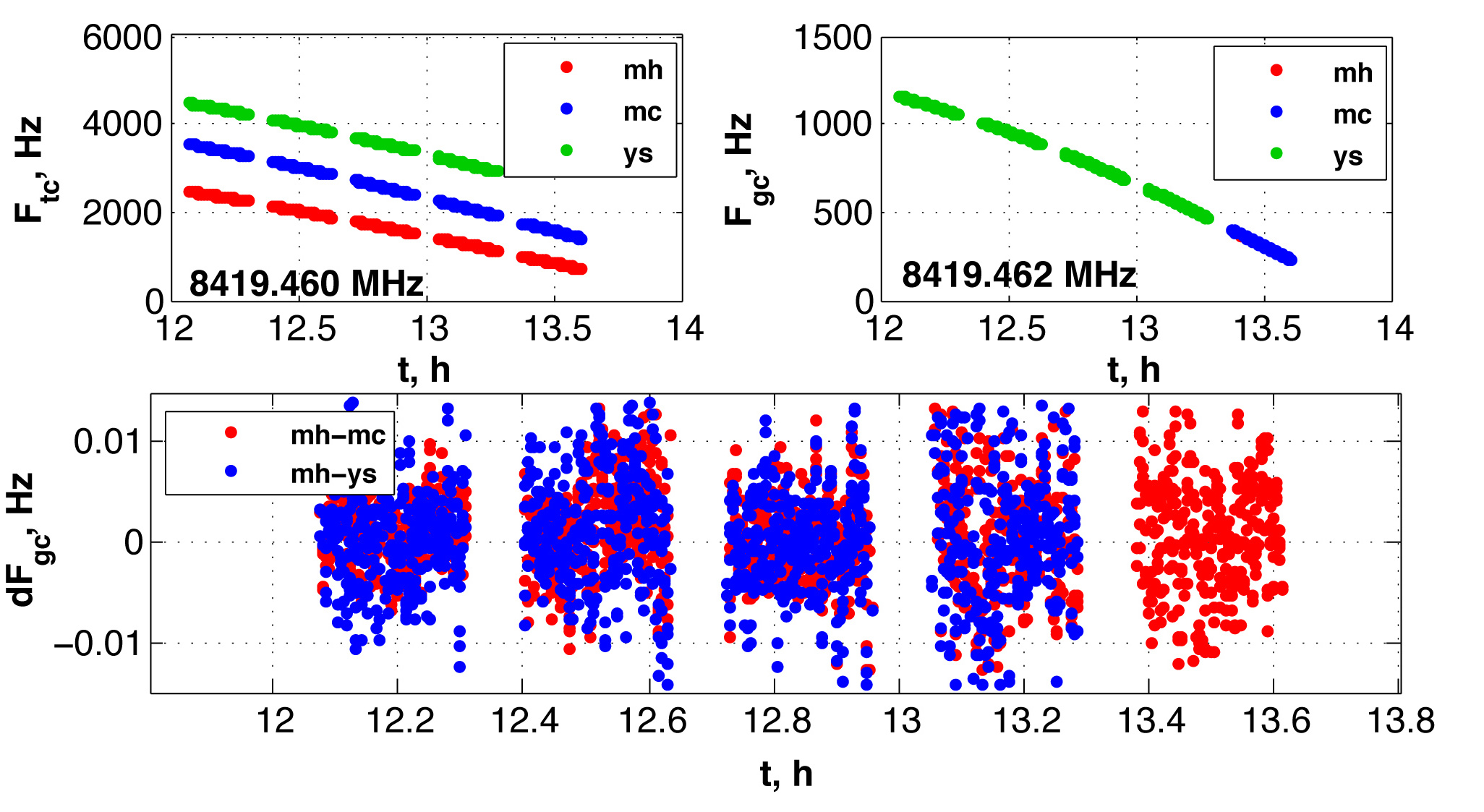}
   \caption{Frequency Doppler shift over the path of the ESA tracking station Cebreros -- VEX -- Mets\"{a}hovi, Medicina, Yebes (top left), Cebreros -- VEX -- Mets\"{a}hovi, Medicina, Yebes - geocentre (top right). The differential frequencies in the geocentre with a standard deviation of$\sim5$~mHz at 2.5 second sampling are shown at the bottom. 2010~August~23, EVN project EM081.}
   \label{fdets}
    \end{figure}

The data were acquired using the standard VLBI data acquisition system based on the Mark5 A/B recording equipment. We used a base frequency of 8411.99~MHz, four 16--MHz--wide channels with 2-bit Nyquist sampling, resulting on aggregate data rate of 256~Mbps per station. The data files were either transferred over the high--capacity network connection after the experiment to Mets\"{a}hovi and JIVE or recorded on disks and then shipped to the processing centre at JIVE.

   \begin{figure}
   \centering
   \includegraphics[width=255pt,clip]{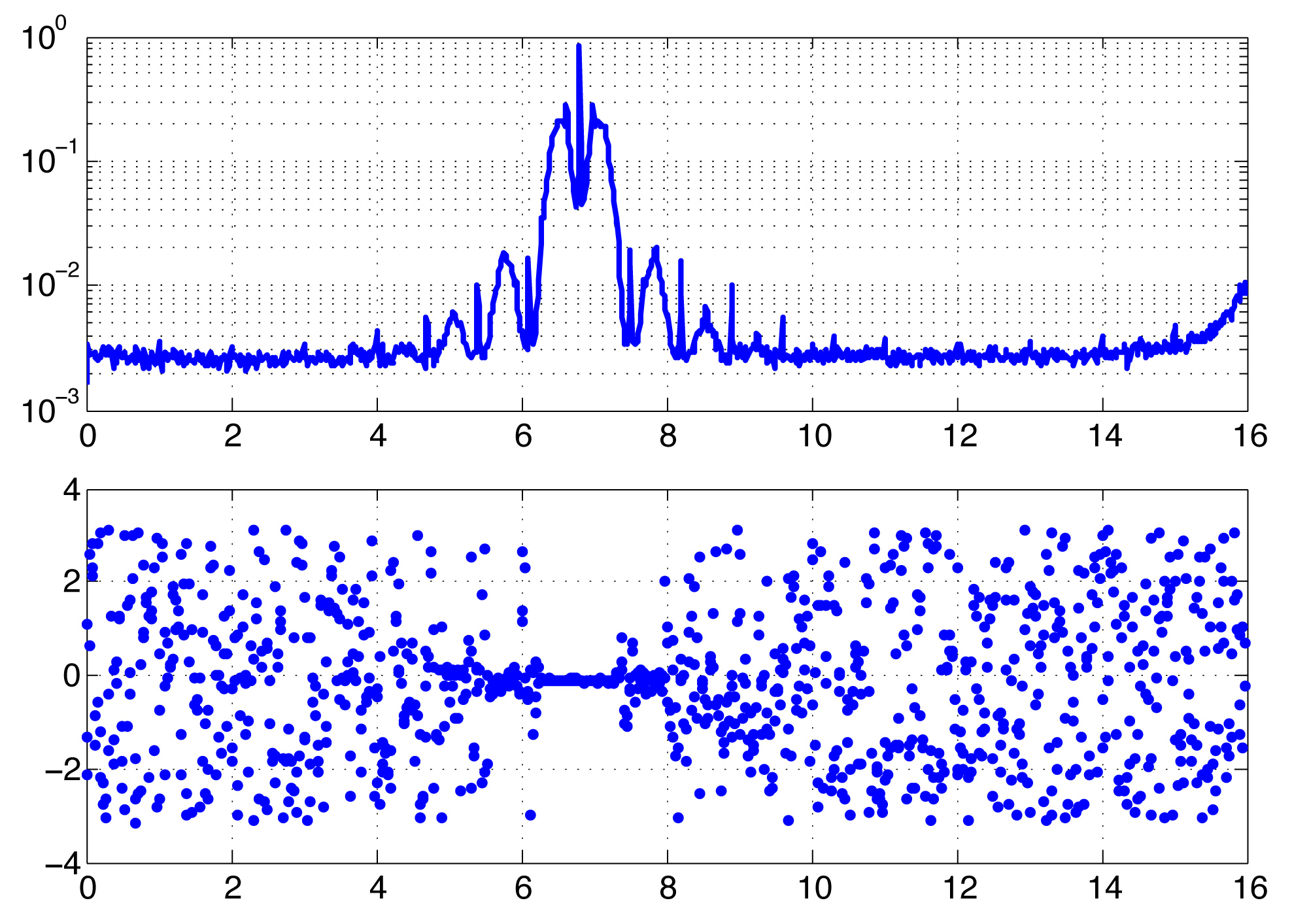}
   \caption{Interferometric fringes of the VEX signal in the spectral domain, 1024 frequency points, baseline Onsala-Zelenchukskaya, 2011 March 28, EVN project EM081, using the near--field delay model and two--way Doppler phase correction to avoid frequency smearing when correlating at the SFXC at JIVE. Top: amplitude in relative units; bottom: phase in radians. The horizontal axis -- frequency in video band (MHz) relative to 8.411990 GHz. Most of the power of the spacecraft signal is concentrated within a $\sim$1 MHz-wide band around the carrier line, although the S/N for the carrier is considerably higher than for the rest of the band.}
   \label{ACheHa}
    \end{figure}

The narrow-band data processing and carrier frequency/phase extraction were conducted with the SWSpec and SCtracker software at Mets\"{a}hovi. The Doppler frequency detections from all stations reduced to the geocentre using the near--field delay model (as $f_\mathrm{gc}(t)=f_\mathrm{tc}(t-\tau)\cdotp(1-\dot{\tau})$) differ on average only by several mHz or even less, proving the consistency between the actual and model delays. As an example, the topocentric frequency detections measured for the stations Mh, Mc and Ys, detections reduced to the geocentre and differential geocentric frequencies for the experiment conducted on 2010 August 23 are shown in Fig.~\ref{fdets}.

   \begin{figure*}
   \centering
   \includegraphics[width=470pt,clip]{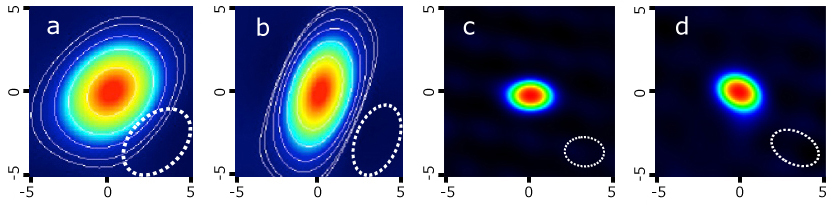}
   \caption{{\bf (a, b)} - Fringe finder (3C446, 2.5 Jy source at an angular distance of $15^{\circ}$ north from the target) and a calibrator (J2211-1328, 0.17 Jy source at an angular distance of $2.5^{\circ}$ south from the target) self-calibrated images using the EVN array excluding Hh, {\bf (c, d)} - self-calibrated images of the same sources obtained with the whole array. Horizontal axis - deviation from the nominal source's Right Ascension in mas, vertical axis - deviation from the nominal source's Declination in mas. Dashed white ellipses at the bottom-right represent the synthesised beam size and shape.}
   \label{calib}
    \end{figure*}
%

   \begin{figure*}
   \centering
   \includegraphics[width=510pt,clip]{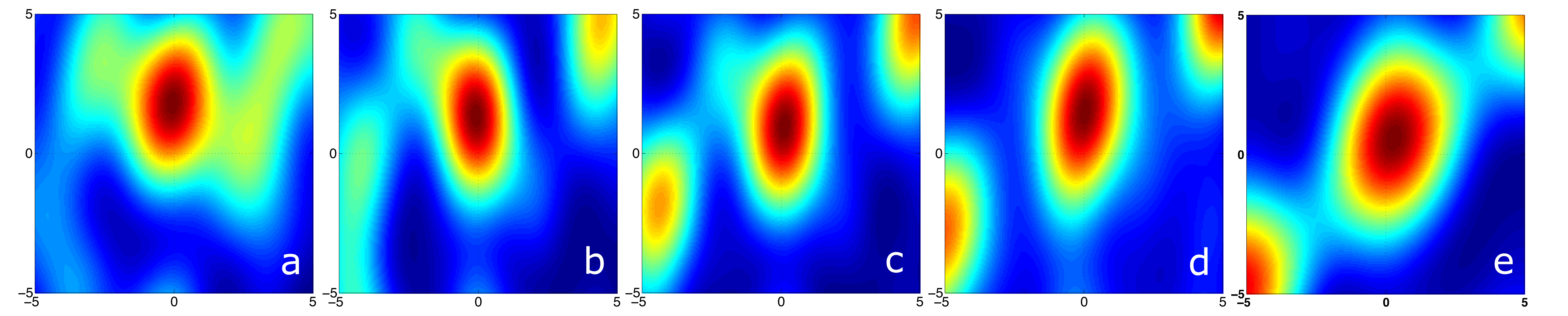}
   \caption{Radio images of VEX obtained with the EVN array. The horizontal axis represents the deviation from the nominal source's Right Ascension in mas, vertical axis -- deviation from the nominal source's Declination in mas. {\bf(a)} 09h05m TDB, stations On, Wz, Ma, Mc, Mh, Sv, Zc, {\bf(b)} 09h30m TDB, Ys joined in, {\bf(c)} 09h55m TDB, Sv left, {\bf(d)} 10h20m TDB, Mh left, {\bf(e)} 10h45m TDB, Zc left. No cleaning applied.}
   \label{image}
    \end{figure*}

The broadband correlation was conducted at JIVE using the SFXC. The fringe finder data were used to perform a clock search. The VEX cross-correlation spectrum is smeared due to the intrinsic change of the frequency emitted by the S/C, as it retransmits the signal in the two-way Doppler regime. This change may reach several kHz over a time span of one hour, which is shown in Fig.~\ref{fdets}. To avoid this smearing when correlated with the SFXC, a two-way Doppler correction (see formula~(\ref{eq1}) and the paragraph below it) was applied to the S/C data. Fig.~\ref{ACheHa} represents the amplitude and phase of the cross-correlation function of the VEX signal (amplitude and phase) on the baseline On--Mh. Most of the spectral power is concentrated within a $\sim$1~MHz region around the carrier line, allowing us to resolve the 2$\mathrm{\pi}$ ambiguity between the group delay detected by the broad--band correlation with the SFXC and phase delay detected using the SCtracker software. Fig.~\ref{calib} (a,b) shows the self-calibrated and cleaned images of the fringe finder and the calibrator obtained with the Difmap \citep{Difmap} and AIPS \citep{AIPS} software.

Unfortunately, an anomalous clock rate offset at Hh did not allow us to obtain its absolute phases. Although good for self-calibrated imaging (see Fig.~\ref{calib}), the Hh data were not used for absolute positioning during the EM081c observational run on 2011 March 28. The Hh LO converter was fixed several days later, based on our analysis of the LO behaviour made with ultra-high spectral resolution SCTrack software package.

   \begin{figure*}
   \centering
   \includegraphics[width=370pt,clip]{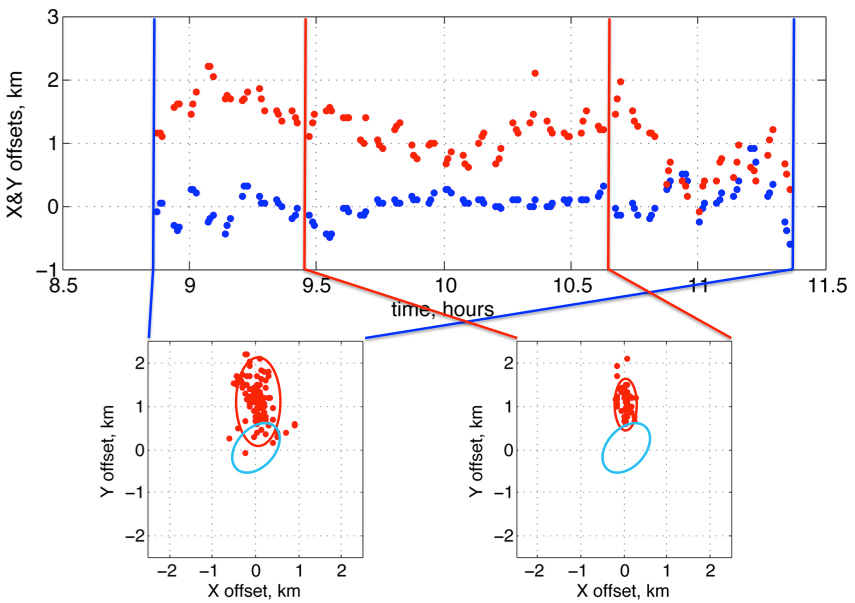}
   \caption{{\bf (top)} Measured coordinate offsets with respect to the a priori trajectory, supplied by ESOC. The X-offset is shown in blue, Y-offset - in red. X and Y are approximately RA and Dec. The measurement sampling -- 30~s, 1~km corresponds to $\approx$ 1.1~mas at the distance of 1.24~AU.
{\bf (bottom)} Error scatter plots; the light blue ellipses show the a priori estimates of the trajectory accuracy, red ellipses -- our measurement's 3$\sigma$ scatter for a full time-range (left) and best-calibrated time range (right)}
   \label{OD}
    \end{figure*}

The correlation of the calibrator source for each baseline provides the residual delays. For coherent phase--referencing, the residual delays must be lower than the radio wave period at X-band, which is about 0.120 ns. The measured residual delays and phases were applied to the detected residual phases of the target. With these we performed imaging of the VEX spacecraft and derived the deviations of the lateral coordinates of the spacecraft with respect to a priori ones. Fig.~\ref{image} shows the imaging results obtained for VEX with the EVN array without any cleaning. The averaging time for each image is 20 minutes. Fig.~\ref{OD} shows the estimated deviations of the a priori lateral position of the VEX spacecraft. The integration time for each point is 30 seconds. The coordinate offsets were estimated using equation \ref{LSQ}. However, for a consistency check, we made S/C images for each short interval of 30 seconds with a ``super"-resolution and measured the peak position on each of them. The results yielded a consistency of the two approaches at the level of 25 microarcseconds. The orbit accuracy estimate is at a 3 sigma level of 200-300 m across the track and 500-600 m along the track.

The north/south VEX offsets of $\sim 1$ mas from the expected orbital position seen in Fig.~\ref{OD}, although nearly within the error budget, may be caused by systematic phase errors between the target and calibrator due to their low elevation. More experiments with different relative positions of the S/C and calibrator(s) are needed to prove or disprove this claim.

\section{Discussion and conclusions}

The EVN experiment EM081 enabled us to show that astrometric spacecraft positioning with a very high accuracy is achievable using the PRIDE approach. A relatively large angular distance between VEX and the calibrator source ($2.5^{\circ}$), their low elevation and a limited accuracy of the troposphere/ionosphere models did not allow us to include the data from St. Croix (Sc) in the astrometric solution without a baseline fringe phase $2\pi$-bifurcation. Fig.~\ref{calib} (c, d) shows the self-calibrated images of the fringe finder and the calibrator with Sc and Hh added. As seen from this figure, more than a three--fold resolution improvement (at the level of up to 20~mas, 0.1~nrad) is possible. We adopt this value as the aim for our future experiments. 
To achieve this increase in resolution, we will have to resolve the $2\pi$ ambiguities for the longer baselines. For this purpose, closer (in-beam in the ideal case), preferably multiple calibrators should be used in the experimental setup for phase referencing. In addition to this, better troposphere and ionosphere models would be very important, especially for the low elevations of the sources. 

The results of the current study in terms of achieved accuracy are compatible with those of the other recent well-established VLBI spacecraft tracking projects, such as the VLBA tracking of the Cassini spacecraft \citep{Jones} and NASA's Mars Exploration Rover B spacecraft during its final cruise phase \citep{Lanyi}.

The PRIDE approach described in this paper proves its applicability to virtually any deep-space mission almost anywhere in the solar system. We note that the approach is not exclusively oriented on planetary science missions, but other deep-space and near--Earth targets. This has been demonstrated by VLBI observations of GLONASS satellites \citep{GNSS10,GNSS11} aiming to establish a better link between the terrestrial and celectial reference frames.

\begin{acknowledgements}
We would like to express our sincere gratitude to T.~Morley (ESA ESOC) for persistent support, M.~P\'{e}rez-Ay\'{u}car (ESA ESAC) for robust assistance and the anonymous referee for useful and constructive suggestions. The EVN is a joint facility of European, Chinese, South African and other radio astronomy institutes funded by their national research councils. The National Radio Astronomy Observatory is a facility of the National Science Foundation operated under cooperative agreement by Associated Universities, Inc. The authors would like to thank the personnel of the participating stations: P.~de Vicente (Yebes), J.~Quick (Hartebeesthoek), G.~Kronschnabl (Wettzell), R.~Haas (Onsala), A.~Orlatti (Medicina), G.~Colucci (Matera), A.~Finkelstein, M.~Kharinov, A.~Mikhailov (Svetloe and Zelenchukskaya of the KVAZAR network) and the staff of the NRAO VLBA.  R. Campbell and A.~Keimpema of JIVE provided important support to various components of the project. We also thank the Royal Netherlands Meteorological Institute (KNMI) for granting access to the ECMWF meteorological data. D.A.~Duev acknowledges the EC FP7 project EuroPlaNet (grant agreement 228319). G.~Cim\'{o} acknowledges the EC FP7 project ESPaCE (grant agreement 263466). T.~Bocanegra Bahamon acknowledges the NWO--ShAO agreement on collaboration in VLBI.

\end{acknowledgements}

\end{document}